\UseRawInputEncoding

\documentclass[]{raa}
\usepackage{graphicx,times}
\usepackage{natbib}
\usepackage{amssymb,amsmath}
\usepackage{graphicx}
\usepackage{epstopdf}
\bibpunct{(}{)}{;}{a}{}{,}
\makeatletter

\newcommand{\Rmnum}[1]{\expandafter\@slowromancap\romannumeral #1@}
\makeatother

\usepackage[a4paper=true,pagebackref=true]{hyperref}
\hypersetup{pdftitle = The title of my PDF, pdfauthor = My name, pdfsubject= The subject, pdfkeywords = keyword1 keyword2 keyword3}
\hypersetup{colorlinks = true, linkcolor = green, anchorcolor = red, citecolor = blue, filecolor = red, pagecolor = red, urlcolor = red}

\begin{document}

   \title{Optical Properties of C$-$rich ($^{12}$C, SiC and FeC) Dust Layered Structure of Massive Stars
$^*$
\footnotetext{\small $*$ Supported by the National Natural Science Foundation of China.}
}

 \volnopage{ {\bf 20XX} Vol.\ {\bf X} No. {\bf XX}, 000--000}
   \setcounter{page}{1}

   \author{Ruiqing Wu\inst{1}, Mengqiu Long\inst{1}, Xiaojiao Zhang\inst{1}, Yunpeng Wang\inst{1}, Mengli Yao\inst{1},
   Mingming Li\inst{1}, Chunhua Zhu\inst{2}, Guoliang L\"{u}\inst{2}, Zhaojun Wang\inst{2},
   Jujia Zhang\inst{3}, Zhao Wang\inst{4}, Wujin Chen\inst{5}.}

%% Here is an example of three authors come from different institutes.
%% For single author or all the authors from an institute, use "\inst{}" only

   \institute{School of Physics and Electronics, Central South University, Changsha 410083, China\inst{1};\\
    School of Physical Science and Technology, Xinjiang University, Urumqi 830046, China\inst{2}; \\
    Yunnan Observatories (YNAO), Chinese Academy of Sciences, Kunming 650216, China\inst{3}; \\
    School of Chemistry and Chemical Engineering, Guangxi University, Nanning 530004, China\inst{4}; \\
    School of Medicine, Xinjiang Medical University, Urumqi 830011, China\inst{5}; \\
   {\it ruiqingwu163@163.com, mqlong@csu.edu.cn}\\
%% Please give the E-mail address of the author, to whom future correspondence and
%% offprint requests will be sent.
\vs \no
   {\small Received 2021 December 27; accepted}
}

\abstract{The composition and structure of interstellar dust are important and complex for
the study of the evolution of stars and the \textbf{interstellar medium} (ISM).
However, there is a lack of corresponding experimental data and model theories.
By theoretical calculations based on ab-initio method, we have predicted and geometry optimized the structures of Carbon-rich (C-rich) dusts,
carbon ($^{12}$C), iron carbide (FeC), silicon carbide (SiC), even silicon ($^{28}$Si), iron ($^{56}$Fe),
and investigated the optical absorption coefficients and emission coefficients
of these materials in 0D (zero$-$dimensional), 1D, and 2D nanostructures.
Comparing the \textbf{nebular spectra} of the supernovae (SN) with the coefficient of dust,
we find that the optical absorption coefficient of the 2D $^{12}$C, $^{28}$Si, $^{56}$Fe, SiC and FeC structure
corresponds to the absorption peak displayed in the infrared band (5$-$8) $\mu$$m$
of the spectrum at 7554 days after the SN1987A explosion.
And it also corresponds to the spectrum of 535 days after the explosion of SN2018bsz,
when the wavelength in the range of (0.2$-$0.8) and (3$-$10) $\mu$$m$.
Nevertheless, 2D SiC and FeC corresponds to the spectrum of 844 days after the explosion of SN2010jl,
when the wavelength is within (0.08$-$10) $\mu$$m$.
Therefore, FeC and SiC may be the second type of dust in SN1987A corresponding to infrared band (5$-$8) $\mu$$m$ of dust
and may be in the ejecta of SN2010jl and SN2018bsz.
The nano$-$scale C$-$rich dust size is $\sim$ 0.1 \emph{nm} in SN2018bsz,
which is 3 orders of magnitude lower than the value of 0.1 $\mu$$m$.
In addition, due to the ionization reaction in the supernova remnant (SNR),
we also calculated the Infrared Radiation (IR) spectrum of dust cations.
We find that the cation of the 2D layered (SiC)$^{2+}$ has a higher IR spectrum than those of the cation (SiC)$^{1+}$ and neutral (SiC)$^{0+}$.
}

   \authorrunning{R.-Q. Wu et al. }            %author_head in even pages
   \titlerunning{Optical Properties of C$-$rich Dust
   }  % title_head in odd pages
   \maketitle

%________________________________________________ sections below
%
\section{Introduction}           %% first-level sections will be auto-capitalized
\label{section1}
Dust plays an important role in the interstellar medium (ISM).
It absorbs optical and ultraviolet (UV) radiation and emit into infrared,
it also affects the spectral energy distribution in the cosmic environment
\citep[e. g.,][]{Gould1963,Cazaux2004}.
Dust is known to form in supernova remnant (SNR) that is rich in heavy elements, in the stellar wind of asymptotic giant branch (AGB) stars,
and in nova ejecta for theoretical research \citep{Zhu2013,Iliadis2018,Bose2019,Zhu2019,Duolikun2019,Gail2020,Lugaro2020}.
Recent years observations indicate that dust can be effectively formed in supernovae (SN) ejecta
\citep{Barlow2010,Dwek2010,Gomez2012,Indebetouw2014,Bevan2016,DeLooze2017,Sarangi2018,Rho2021},
a large number of observation strategies have obtained the high dust mass between $\sim 0.1 M_{\odot}$ and $1 M_{\odot}$
\citep{Dunne2009,Sibthorpe2010,DeLooze2017,Bevan2017,Priestley2019}.
The dust contribution of massive stars to ISM mainly includes silicate and Carbon$-$rich (C$-$rich) dust,
it include heavy elements $^{12}$C, $^{14}$N, $^{16}$O, $^{24}$Mg, $^{26}$Al, $^{28}$Si, $^{56}$Fe.
Especially $^{12}$C, $^{28}$Si and $^{56}$Fe are more abundant,
which heavy$-$element abundance can reach $\sim 10^{-4}$-$10^{-2}$.
The initial mass range of massive stars is [10-30]$M_{\odot}$, and the metallicity is [Fe/H=-2],
the abundance of the heavy elements $^{12}$C, $^{14}$Si, and $^{56}$Fe is in the range $\sim$ [0.028-0.9], [0.001-0.006], [0.0001-0.002], respectively.
And the corresponding output of the ejecta reaches [0.1-0.7], [0.09-0.15], and [0.009-0.05] $M_{\odot}$, respectively
\citep{Ercolano2007,Draine2009,Zhang2016,Marassi2019,Wu2021}.
Moreover, iron can be used in the study of dust growth in ISM, which would be consumed more heavily in warm neutral media (WNM),
iron is the opposite of silicon \citep{Zhukovska2016}.
Although the signature spectral feature of SiC has not been detected in supernovae to date,
\cite{Deneault2017} have present a kinetic model of the formation of silicon carbide (SiC) in the expanding and cooling outflows of Type \Rmnum{2} supernova ejecta, and found that when the C/Si ratio is close to unity, 0.1 and 10, the abundance of SiC is between $10^{-9}$ and $10^{-11}$.
Furthermore, \cite{Kodolányi2018} used transmission electron microscopy (TEM) to study micron-sized (0.99$\times$0.66-2.23$\times$1.64 $\mu$$m^{2}$) SiC from the SN dusts,
and there are enough heavy elements $^{12}$C, $^{28}$Si to condense into SiC dust in the SNR.
In addition, the sub-grain with high $^{56}$Fe content and large $^{28}$Si loss may be $^{56}$Fe metal,
while the sub-grain with $^{28}$Si enrichment or small $^{28}$Si loss may be $^{56}$Fe silicide \citep{Singerling2021}.
Here, we believe that this type of $^{56}$Fe metal tends to be $^{56}$Fe and FeC.

The formation of carbides (e.g., SiC, Fe$_{3}$C) can be predicted by thermodynamic equilibrium calculation
for given temperature, pressure, and C/O ratio in the circumstellar materials (CSMs),
including AGB star and Core-collapse supernovae (CCSNe) \citep{Lodders1995,Sharp1995,Hoppe2001,Singerling2021}.
The size, shape and corresponding energetics of the dust grains are indispensable information for these calculations.
For the hypothetical astronomical silicon dust temperature is higher than the condensation temperature
$\sim$ 1000 $K$ \citep{Nozawa2003}.
\cite{Gail1999} studied the limit of the dust condensation zone temperature (900$-$1300 $K$)
corresponds to the sublimation temperature outflow at low pressure ($10^{-9}-10^{-12} bar$).
In order to keep the problem simple,
\cite{Mathis1977} assumed that the particles are spherical and have a size distribution of (0.005$-$0.25) $\mu m$.
\cite{Nozawa2003} showed that the newly formed dust is mainly graphite or amorphous carbon.
In addition, when the radius is $\lesssim$ 0.1 $\mu$$m$,
the condensation temperature of C-rich dust is $\sim$ (1300$-$1350) $K$,
and the formation temperature of Fe$_{3}$C is between 1200 $K$ and 1500 $K$ \citep{Singerling2021}.

\cite{Dwek2010} found via Spitzer satellite that silicate dust are best suited for the evolution of (8$-$30) $\mu$$m$ spectrum and time data.
The Infrared Radiation (IR) spectrum (6000$-$8000 day) also shows the presence of very small and thermal (\emph{T} $\sim 350 K$) secondary dust,
these dust are likely to coexist with silicate in the SNR.
Until now, the origin of this emissions component is an unsolved problem.
The temperature of the secondary dust component is significantly higher than that of the silicate grains,
ranging from 370 $K$ (iron dust) to 460 $K$ (C-rich dust).
Therefore, FeC and SiC dust are likely to be present in the SNR.
According to the above description,
we ignored the temperature and pressure, because these factors are very low.

Here, we simulate the nano$-$layered structure of FeC and SiC in 0D, 1D and 2D in the SNR,
the structures of dust including metallic needles \citep{Dwek2004},
multi$-$layered particles \citep{Voshchinnikov2017}, hydrogenated nanotubes (HNT) molecular growth \citep{Chen2019},
spherical \citep{Fischera2004,Demyk2017,Gail2020}.
\cite{Marassi2019} and \cite{Kirchschlager2020} groups have studied the dust mass in the SNR,
furthermore, some groups research the spectral of the iron and silicates dust \citep{Maldoni2005,Gail2020}.

The radiation output of SN 1987A is mainly by the radioactive decay energy in the ejecta,
which is affected by the interaction of its shock wave with the inner equatorial ring (ER).
The ER may be formed by mass loss during the evolution of stars \citep{Heger1998}.
We do not consider the factors affecting dust radiation, because the mechanism has not been perfected and recognized.
Noted that the total mass of dust particles ($M_{d}$), $κ_{ν}$ and temperature ($T_{d}$)
are positively correlated with the dust heat flux emission $F_{ν}$,
and the dust mass absorption coefficient ($\kappa_{\nu}$) depends on the dust type and size distribution \citep{Maeda2013}.
We thus consider the \textbf{absorption coefficient} \textbf{($\alpha$)} and the \textbf{emission coefficient} \textbf{(E)} of the C$-$rich dust in the SNR,
and estimated the radiant flux of C-rich dust.

In SN 2010jl, dust formation can only commence after day $\sim380$ \citep{Sarangi2018}.
\cite{Chen2021} propose that Super Luminous Supernova (SLSN) 2018bsz can form new dust in the ejecta more than 200 days after the SN explosion.
SN 2018bsz is the first to prove the presence of dust in SLSN ejecta.
The carbon dust output can reach 10$^{-2}$ M$_\odot$ in 535 days after the SN explosion.
The SN forward shock wave is sufficient to heat and ionize the CSM,
which will generate a very strong ionizing radiation flux \citep{Sarangi2018}.
The energy of the interaction between the CSM and the shock wave depends on the SN photosphere \citep{Fransson2014}.
The pre$-$existing dust in the CSM can be formed in cooling gas after the post-shock.
According to photo$-$ionization by the central star, \cite{Ota2019}
predicts the vibrational spectra of ionized graphene and fullerene (C$_{60}$),
and their results are in reasonable agreement with laboratory experiments.

\cite{Chen2019} studied the IR vibrational spectra of HNT and their cations,
\cite{Devi2020}, \cite{Zhang2020}, and \cite{Hanine2020} researched the IR spectra of five and six membered ring PAH, respectively.
However, few research groups use first principles to study the correlation between the IR spectrum
of the different structure of C$-$rich dust and the SN spectrum.
Through this method, we can obtain more accurate and reliable dust structure, composition and size,
and better study the mass of dust, interstellar extinction from the CSM \citep{Schlafly2011,Dhar2017,Chen2021}.
Therefore, it is necessary to study whether the C$-$rich dust structures of different dimensions are related to the spectrum of SNR.
This is also conducive to later observations on the structure and shape of dust.

The physical environment for the formation of dust is a very complex and difficult content.
In this paper, we ignore some of the factors and focus on the influence of dust structure and composition on the spectrum.
In Section 2, the relevant physical parameters of the dust model are mainly explained.
The detailed results are discussed in Section 3.
Section 4 is the main conclusion.

\section{Model}\label{sec:2}

We use the open$-$source CALYPSO (Crystal structure AnaLYsis by Particle Swarm Optimization, version 6.0)
software to acquire the initial structure of C$-$rich dust \citep{Wang2010,Su2017,Yin2020}.
Then we predict stable or metastable nanostructures based on the chemical composition given by CALYPSO interfaced
with Vienna Ab$-$initio Simulation Package (VASP).

Parameters for 2D structure prediction (2D=T).
In addition, the Ialgo is 2 for local Particle Swarm Optimization (PSO).
Local optimization increases the cost of each structure, but it effectively reduces efficiency,
enhances the comparability between different structures, and provides a local optimal structure for use.
PSO algorithm can predict reliable ground state and metastable structures,
generated symmetry structure, and then the elimination of similar structures to obtain the initial dusty geometric structure.
When the temperature \emph{T}=0 K, we take the enthalpy of the free energy reduced to the locally optimized structure
as the fitness function of the calculation simulation. The atomic coordinates and lattice parameters are locally optimized,
and a large number of the worst structures are eliminated.
\textbf{The structure below has entered the next$-$generation structure prediction through PSO} \citep{Wang2010}.
\textbf{We predicted} \textbf{(FeC)$^{0+}$ and (SiC)$^{0+}$ } (\textbf{4 formula unit per cel})
\textbf{structures with the only input information of chemical composition}.
Population size is set to 30 \citep{Luo2011,Wang2010}.
The distance between two adjacent layers is 2 (Layer Gap=2 {\AA}).
The total energy is calculated from the beginning with 0 K and 0 GPa, so the free energy is reduced to enthalpy \citep{Wang2010}.
In the PSO local search, the structure in the phase space can be regarded as a particle,
the structure composed of a group of particles is called a population or generation,
and the particle position is mainly updated by the following equation:
\begin{equation}
x_{\rm i,j}^{\rm t+1} = x_{\rm i,j}^{\rm t} + \nu_{\rm i,j}^{\rm t+1},
\label{yields}
\end{equation}
where, \emph{x} and $\nu$ represent position and velocity of an atom, respectively,
i is the exponent of the atom, j is the dimension of the structure, and t is the generation index \citep{Luo2011}.

Moreover, the parameter settings for VASP are as follows: the energy convergence of standard is $1 \times 10^{-6}$ eV,
the Hellmann-Feynman forces of standard is $\leq 1 \times 10^{-2}$ eV/{\AA}, the vacuum layer is 20 {\AA} along the z direction \citep{Zhang2018,Yao2020}.
The unit cell of FeC/SiC crystal is in the Figure~\ref{Fig1} subgraph (a), (e), and (i),
and the optimized lattice constant are calculated as a=b=c=3.382 \AA,  a=b=c=2.611 \AA, and a=b=c=4.379 \AA, respectively.
Lattice angle are $\alpha$=$\beta$=$\gamma$=90 degree.
Here, \textbf{the vacuum layer is order to prevent the dust structure from reacting with other units structural}.
In the PSO simulations, via the density functional theory (DFT) based on the plane$-$wave calculations,
we optimized the structures and calculated the related physical quantities \citep{Luo2011}.
The k-point grid in the Brillouin zone is set to $1 \times 1 \times 1$ for static calculation.
The calculation is based on the iterative method of Kohn-Sham, and the plane wave of the projected wave
pseudopotential is concentrated on the DFT equation \citep{Kohn1965}.
In this paper, the calculation function is Perdew-Burke-Ernzerhof (PBE), the generalized gradient approximation function of exchange correlation,
and the cut$-$off energy of the plane wave is 400 eV \citep{Rani2014,Lin2016,Yao2020}.
Two dimensional structures are constructed by 5 $\times$ 5 supercell for 32(FeC)$^{0+}$ (Consisting of 32 atoms is electrically neutral dust),
and 4 $\times$ 4 supercells for 85(SiC)$^{0+}$, and 72(Si)$^{0+}$.
The linear optical properties of dust can be obtained from the frequency-dependent complex dielectric function:
\begin{equation}
\varepsilon(\omega) = \varepsilon_{1}(\omega) + i\varepsilon_{2}(\omega),
\label{yields}
\end{equation}
where $\varepsilon_{1}(\omega)$ and $\varepsilon_{2}(\omega)$ are the real and imaginary parts of the dielectric function,
and $\omega$ is the photon frequency.
Absorption coefficient $\alpha(\omega)$ is:
\begin{equation}
\alpha(\omega) = \frac{\sqrt{2}\omega}{c} [\sqrt{\varepsilon_{1}(\omega)^{2} + \varepsilon_{2}(\omega)^{2}}- \varepsilon_{1}]^{\frac{1}{2}},
\label{yields}
\end{equation}
here, \emph{c} is the speed of light.
The reflectivity $R(\omega)$ is calculated by this formula:
\begin{equation}
R(\omega) = \frac{(n-1)^{2} + K^{2}}{(n+1)^{2} + K^{2}},
\label{yields}
\end{equation}
where \emph{n} and \emph{K} are the real and imaginary parts of the refractive index, respectively \citep{Wang2019,Yao2020,Rani2014}.
Regarding the ionization state of dust,
the total number of electrons is set to randomly lose one or two electrons in VASP.

In the FeC model, the optimized Fe-C bond length is (1.82-1.86) \AA, it is close to the value in the FeC$_{2}$ structure (1.84-2.11) \AA \citep{Zhao2016}.
For the SiC model, the optimized Si-C bond length is (1.89-1.92) \AA, which is 1.8 \AA in the work of \cite{Zheng2011}.
Moreover, we calculated the stability of the cohesive energy verification structural model,
through the formula:
\begin{equation}
E_{coh}=(nE_{fe/si}+mE_{c}-E_{tol})/(n+m),
\label{yields}
\end{equation}
where, $E_{fe/si}$, $E_{c}$, and $E_{tol}$ are the energy of a single$^{56}$Fe/$^{28}$Si atom, $^{12}$C atom and total energy of the monolayer FeC/SiC structure, respectively.
These n and m are the number of $^{56}$Fe/$^{28}$Si and $^{12}$C atoms in the monolayer structure, respectively.
The calculated cohesive energy of per atom of FeC and SiC is 7.13 eV and 6.61 eV, respectively.
The positive values indicate that the cohesive process is exothermic, and which are larger than the cohesive energy 5.76 eV per atom of tetragonal-FeC (t-FeC) and 5.59 eV per atom of orthorhombic-FeC (o-FeC),
Be$_{5}$C$_{2}$ (per atom 4.58 eV), and Be$_{2}$C (per atom 4.86 eV) \citep{Li2014,Wang2016,Fan2020}.
Therefore, such a high cohesion can keep the FeC and SiC monolayer structures as a stability connection network.

\section{Results}\label{sec:3}

We calculated the optical coefficients of different structures of
$^{56}$Fe, $^{28}$Si, $^{12}$C, FeC, and SiC from nanoparticles,
in order to discuss the contribution of the structure and composition of dust spectrum of massive stars.
Specifically, we compared and analyzed the light absorption coefficient and emission coefficient of different types of dust,
and the IR spectrum with the observation data in the work of \cite{Dwek2010} (SN 1987A),
\cite{Sarangi2018} (SN 2010jl), and \cite{Chen2021} (SN 2018bsz).
The IR spectrum with ionization conditions of the C$-$rich dust are consistent with the observed
and theoretical values within a certain wavelength range.

\subsection{IR spectrum of C$-$rich dust}\label{sec:3.1}

\begin{figure}[ht!]%"[]"中为位置参数，四个参数tbph依次是置顶、置底、浮动、当前位置，，选用的参数优先顺序为h-t-b-p
\centering
\includegraphics[scale=0.82, angle=0]{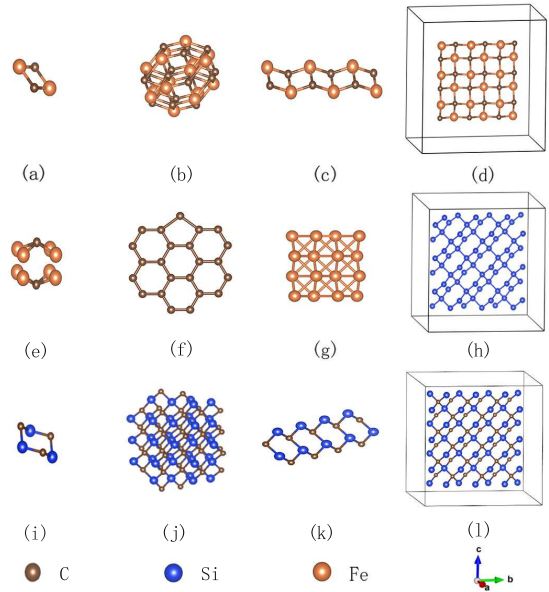}%"scale"后的数字为图形的宽度，也可用"width=1.0\columnwidth"定义
\caption{The initial structure of (FeC)$^{0+}$ dust is (a) model, based on (a) structure,
we have established zero-dimensional (0D) to two-dimensional (2D) structures (See the Figure~\ref{Fig1} for explanation).} % 图题
\label{Fig1}%{}中"fig:example1"为图名，引用时用\ref{fig:example1}
\end{figure}

Figure~\ref{Fig1} shows the predicted structures of 0D 32(FeC)$^{0+}$-Z in (b) structure model,
1D 12(FeC)$^{0+}$-O in (c) structure model and 2D 36(FeC)$^{0+}$-T in (d) structure model.
(e) model is 10(Fe$_{4}$C)$^{0+}$ structure, (f) model is 23(C)$^{0+}$-T structure, (g) and (h) models are 16(Fe)$^{0+}$ and 72(Si)$^{0+}$, respectively.
The initial model of SiC is (i) model, (j), (k), (l) models respectively represent a 0D 91(SiC)$^{0+}$-Z structure,
a 1D 18(SiC)$^{0+}$-O structure and a 2D 85(SiC)$^{0+}$-T structure.
In the legend below: $^{12}$C, $^{28}$Si, $^{56}$Fe atoms are shown as brown, blue, and orange colors, respectively.
We choose the initial models FeC and Fe$_{4}$C,
which are generated in the second generation [(a) model],
seventh generation [(e) model] in the left Figure~\ref{Fig1}, respectively.
And SiC is generated in the sixteenth generations [(i) model].
23(C)$^{0+}$-T (f) model comes from the work of \cite{Ota2019},
\cite{Otsuka2016} indicated that small carbon clusters, small graphite flakes or
fullerenes have the ability to form very small clusters.
According to astronomically observed, the infrared spectrum (IR) in the planetary nebula Tc 1 coincides with the neutral C$_{60}$ vibration spectrum
\citep{Kroto1988,Cami2010,Otsuka2016,Ota2019}.
The 16(Fe)$^{0+}$-T (g) and 72(Si)$^{0+}$-T (h) models are built with Virtual NanoLab with Atomistix ToolKit (VNL$-$ATK, version 2017) \citep{Liu2018,Yi2018}.

\begin{figure}[ht!]%"[]"中为位置参数，四个参数tbph依次是置顶、置底、浮动、当前位置，，选用的参数优先顺序为h-t-b-p
\centering
\includegraphics[scale=0.76, angle=0]{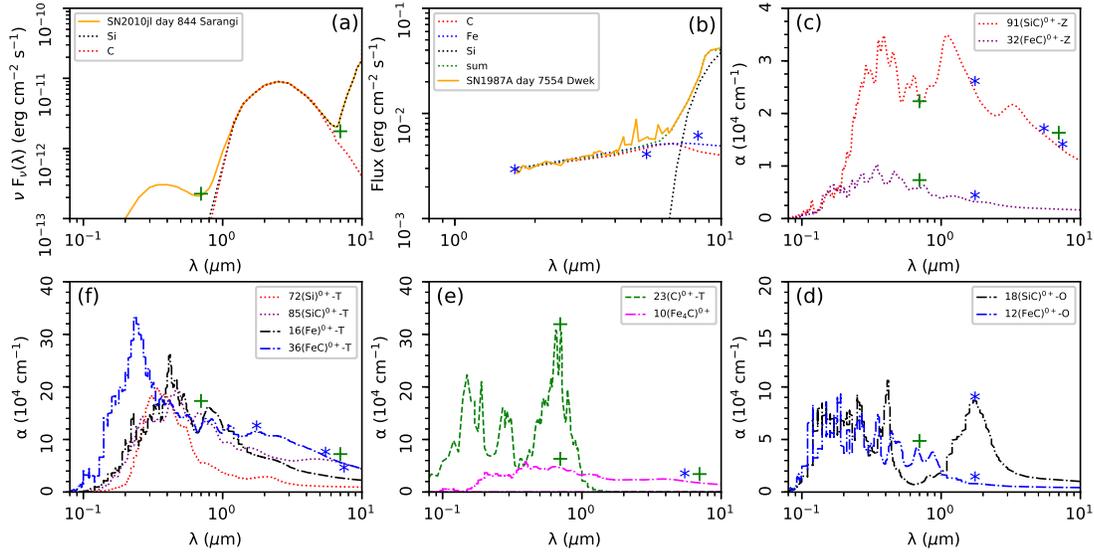}%"scale"后的数字为图形的宽度，也可用"width=1.0\columnwidth"定义
\caption{Panels (a), (b) are from the work of \cite{Sarangi2018} and \cite{Dwek2010}.
Y-axis: flux is the radiant flux of the star or dust. X-axis: $\lambda$ ($\mu$$m$) is the infrared radiation (IR) wavelength.
In the legend of panel (c), the red color dotted-line represents the light absorption spectrum of 91(SiC)$^{0+}$-Z with 0D structures.
Here, $\alpha$ on the Y-axis represents the absorption coefficient.
In our work, we find that the light absorption peaks of the panels (c), (d), (e), (f) in C$-$rich dust matches the peak of
\cite{Sarangi2018} radiation spectra (green plus sign), and \cite{Dwek2010} (blue star symbol), respectively.
}%图题
\label{Fig2}%{}中"fig:example1"为图名，引用时用\ref{fig:example1}
\end{figure}

The panel (a) of Figure~\ref{Fig2} compares the IR spectrum of SN2010jl
after the explosion of SN 844 days with that of silicate (black dotted line) and carbon (red dotted line) dust.
C-rich dust in the near-infrared and mid-infrared bands $\sim$ (0.08$-$10) $\mu$$m$ fits well with observations, while silicate dust
matches well in wavelengths between 1 $\mu$$m$ and 10 $\mu$$m$.
The orange curve in the panel (b) is taken from the work of \cite{Dwek2010}.
Their SN1987A IR spectrum 7554 days after the SN explosion is compared with the total IR spectrum of silicate (black dotted line),
C-rich dust (red dotted line), iron (blue dotted line), silicate and the second type of dust (green dotted line).
Obviously, the IR spectrum of ER in a single component in dust can fit well with the 180 K IR spectrum and the astronomically spectrum of silicate,
especially at wavelengths over (5$-$8) $\mu$$m$.
It is also shown that the IR spectrum of the single component $^{56}$Fe and $^{12}$C dust, and fit well with wavelengths less (5$-$8) $\mu$$m$.
Nevertherless, in the wavelengths of (5$-$8) $\mu$$m$, we find that \textbf{there is a small gap between the IR spectrum of SN1987A and dust}.
The absorption spectrum of our 85(SiC)$^{0+}$-T is basically consistent with that of 3C-SiC that of \cite{Fan2018} at picks of wavelength 0.1, 0.15, 0.2 $\mu$$m$.
\cite{Dwek2010} proposed that there is a second form of dust.
As we can see, the absorption coefficient is inversely proportional to the radiation intensity.
This is consistent with the thermal radiation relation proposed by \cite{Kirchhoff1860},
\begin{equation}
\frac{E}{\alpha} = e,
\label{yields}
\end{equation}
where, \emph{E} is the emission coefficient, $\alpha$ is the absorption coefficient, and \emph{e} is equal to:
\begin{equation}
e = I\frac{\omega_{1}\omega_{2}}{s^{2}},
\label{yields}
\end{equation}
where, \emph{I} is just a function of wavelength and temperature \emph{ I ($\lambda$,T)},
the value of \emph{e} depends on the shape and relative position of the \textbf{openings} \emph{\textbf{$\omega_{1}$}} \textbf{and} \emph{\textbf{$\omega_{2}$}},
which refer to the projection of the opening on the plane perpendicular to the observation axis.
\emph{s} is the distance of the opening.
Furthermore, according to the works of \cite{You1998}, \cite{Rybicki1979}, and \cite{Shu1991},
when the optical depth is close to infinitely small, then the above formula is satisfied (\emph{F} is the flux of the dust):
\begin{equation}
\frac{E}{\alpha} = F,
\label{yields}
\end{equation}

\begin{figure}[ht!]%"[]"中为位置参数，四个参数tbph依次是置顶、置底、浮动、当前位置，，选用的参数优先顺序为h-t-b-p
\centering
\includegraphics[scale=0.76, angle=0]{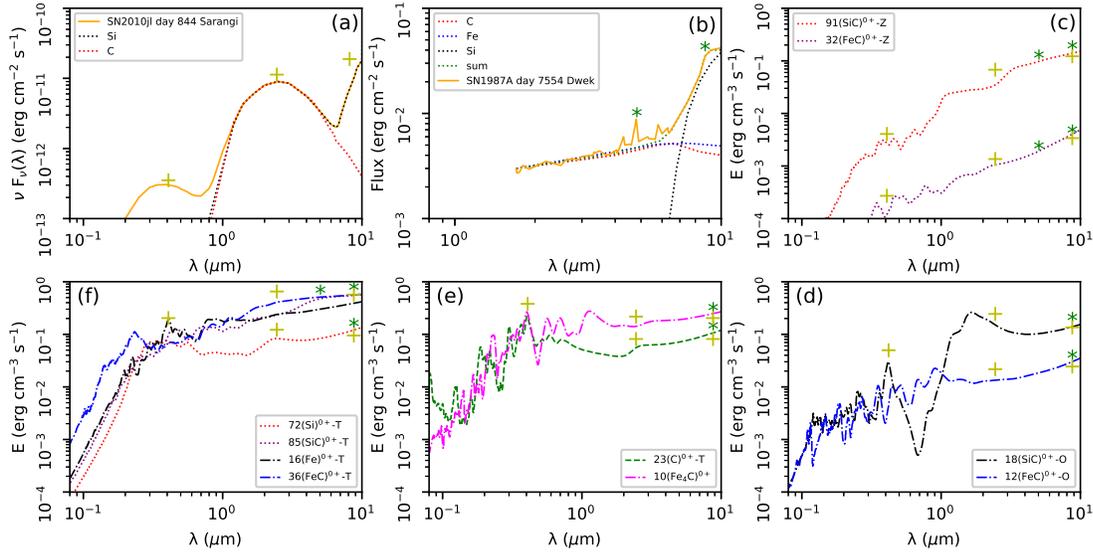}%"scale"后的数字为图形的宽度，也可用"width=1.0\columnwidth"定义
\caption{Similar to the Figure~\ref{Fig2}, we show that the peaks of the light emission spectrum of
(c), (d), (e), (f) (0D-2D) C$-$rich dust are coincide with the IR spectra of \cite{Sarangi2018}
(yellow-green plus sign), and \cite{Dwek2010} (green star symbol).
Here, E on the Y-axis represents the emission coefficient.} %图题
\label{Fig3}%{}中"fig:example1"为图名，引用时用\ref{fig:example1}
\end{figure}

Figure~\ref{Fig3} shows the correlation between emission coefficient of our C$-$rich dust and the SN radiation flux.
Obviously, the correlation satisfies the proportionality suggested by \textbf{equation (8)}.
As indicated by the peaks at the wavelength of 0.4, 2.5, 10 $\mu$$m$.
While light shines on a smoother material surface, the reflected light will be stronger,
then the reflectivity will increase to (0.8$-$0.9).
The photons emitted from the SN cause light echoes by scattering \citep{Yang2017},
the light echoes is the radiation phenomenon formed by the reflection of the stellar light on the surface of the dust.
\cite{Fan2018} believe that as the stellar light travels longer,
the SN radiant flux and echo are both higher at shorter wavelengths.
Here, we make the reflectivity of the dust approximately equal to the emissivity of the dust.
Our 85(SiC)$^{0+}$-T emission spectrum is also basically consistent with the 3C-SiC emission spectrum of \cite{Fan2018}
at the peaks of wavelength 0.14, 0.15, 0.25 $\mu$$m$.

\begin{figure}[ht!]%"[]"中为位置参数，四个参数tbph依次是置顶、置底、浮动、当前位置，，选用的参数优先顺序为h-t-b-p
\centering
\includegraphics[scale=0.76, angle=0]{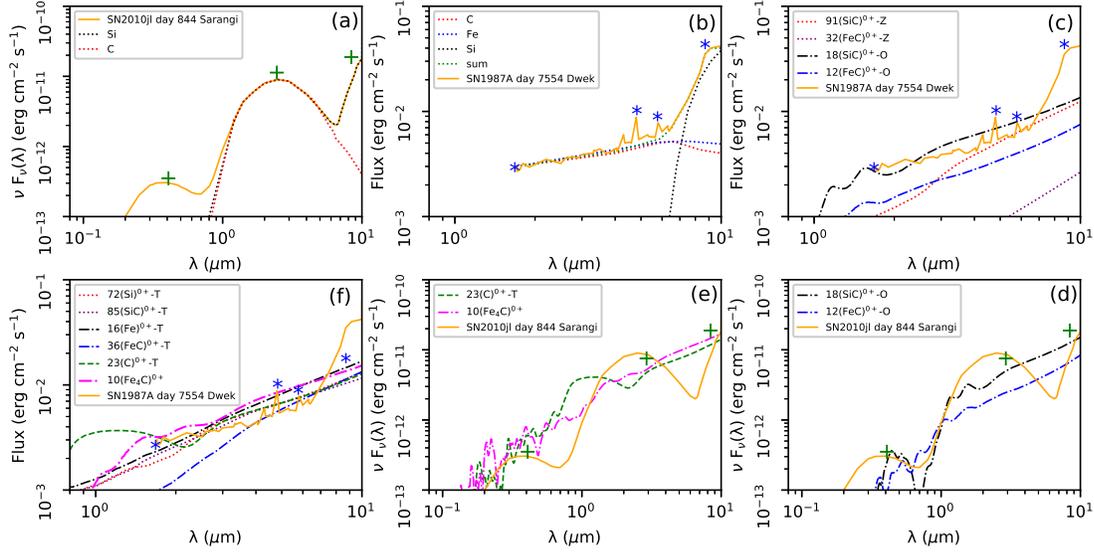}%"scale"后的数字为图形的宽度，也可用"width=1.0\columnwidth" 定义
\caption{The IR spectrum of C-rich dust in different dimensions is compared with the spectrum observed SN1987A and SN2010jl.
It can be seen that they are in good agreement. (c)$-$(f) subgraphs are our work.} % 图题
\label{Fig4}%{}中"fig:example1"为图名，引用时用\ref{fig:example1}
\end{figure}

Figure~\ref{Fig4} shows that the SN spectrum and the C$-$rich dust spectrum are in good agreement, especially with the SN1987A spectrum.
There is a good overlap at the wavelength of (5-8) $\mu$$m$, such as the wavelength range corresponding to the blue star symbol.
Therefore, there is likely to be a 2D or 1D structures of (SiC)$^{0+}$ or (FeC)$^{0+}$ in the remnant of SN1987A.
In particular, the spectrum of the 2D structure is more compatible with the observation than the spectrum of the
0D and 1D structures.
However, 32(FeC)$^{0+}$ has a higher error in its 0D structure.
This may be attributed to the small size of dust clusters in our calculations.
In the panels (c) and (f) of Figure~\ref{Fig4}, \textbf{referring to} the flux of SN1987A,
we calculate the radiant flux of dust based on equation (8), that the size of dust clusters in the SNR is 900 times of C$-$rich dust,
the initial length range is (0.1-0.8) ${\rm \AA}$, the size of the dust close to (9-72) $nm$,
which is exactly in the range (4.5 ${\rm \AA}$$-$0.3 $\mu$$m$) of dust size given by \cite{Draine2021} and \cite{Lugaro2020}.
The radius of the star-shaped silicate sphere and carbonaceous sphere with a wavelength range
from $\lambda$ = 0.1 $\mu$$m$ to $\lambda$ = 4 $\mu$$m$ is about 0.0005 $\mu$$m$ to 0.5 $\mu$$m$ \citep{Draine2011},
and dust \emph{r} $\sim 0.1 \mu$$m$ \citep{Todini2001}.
In the panels (d) and (e) of Figure~\ref{Fig4}, in the flux of SN2010jl,
because the flux of SN2010jl has \textbf{$\nu$Flux}, then again,
we assumed that the frequency of dust per unit time is 1.11 $\times 10^{-9}$ times of flux.
Obviously, the 2D layered structure dust and the small dust cluster 10(Fe$_{4}$C)$^{0+}$ are in good agreement with
the SN1987A spectrum in the wavelength range of (2-8) $\mu$$m$,
followed by the 0D and 1D (SiC)$^{0+}$ and (FeC)$^{0+}$ structures.
In addition, in Figure~\ref{Fig4} (d), the 1D (FeC)$^{0+}$ and (SiC)$^{0+}$ structures in the wavelength range of (0.4-1.2) $\mu$$m$,
with peaks of 3.5, 4.5, 8, 9 $\mu$$m$, and SN2010jl spectra are in good agreement.
In the sub$-$figure Figure~\ref{Fig4} (e), 23(C)$^{0+}$-T and 10(Fe$_{4}$C)$^{0+}$ are fit well with SN2010jl spectral in the wavelength
range of (0.2-0.5) $\mu$$m$ and the peak values are 1, 3, 10$\mu$$m$.
In the sub$-$figure Figure~\ref{Fig4} (f), 23(C)$^{0+}$, 16(Fe)$^{0+}$, 85(SiC)$^{0+}$,
and 72(Si)$^{0+}$ 2D models and the spectrum of SN1987A [sub$-$figure (b)] are in good agreement in the wavelength range of (1.6-8.0) $\mu$$m$.
There are still gaps in some data, which may be caused by errors in the parameters of our model,
for example, we have ignored the temperature, pressure, etc.

\subsection{IR spectrum of C$-$rich dust in ionized state}\label{sec:3.2}

In diffuse ISM, the density of UV photons is \textbf{proportional} to the electronic density of states,
but the speed of photons is always much faster than electrons.
The photoionization of the central star will cause a large number of escaped photons.
If the rate of photon absorption is greater than that of electrons colliding with particles,
photoelectric charging can drive particles to a positive potential,
thus the particles are electropositive.
If there is not enough energy to overcome the potential of positively charged particles,
it will result in Coulomb focusing increases in the electrons collision of crystal dust \citep{Draine2011}.
Therefore, we also study and analyze the spectral distribution of C-rich dust in the ionized state.

\begin{figure}[ht!]%"[]"中为位置参数，四个参数tbph依次是置顶、置底、浮动、当前位置，，选用的参数优先顺序为h-t-b-p
\centering
\includegraphics[scale=0.76, angle=0]{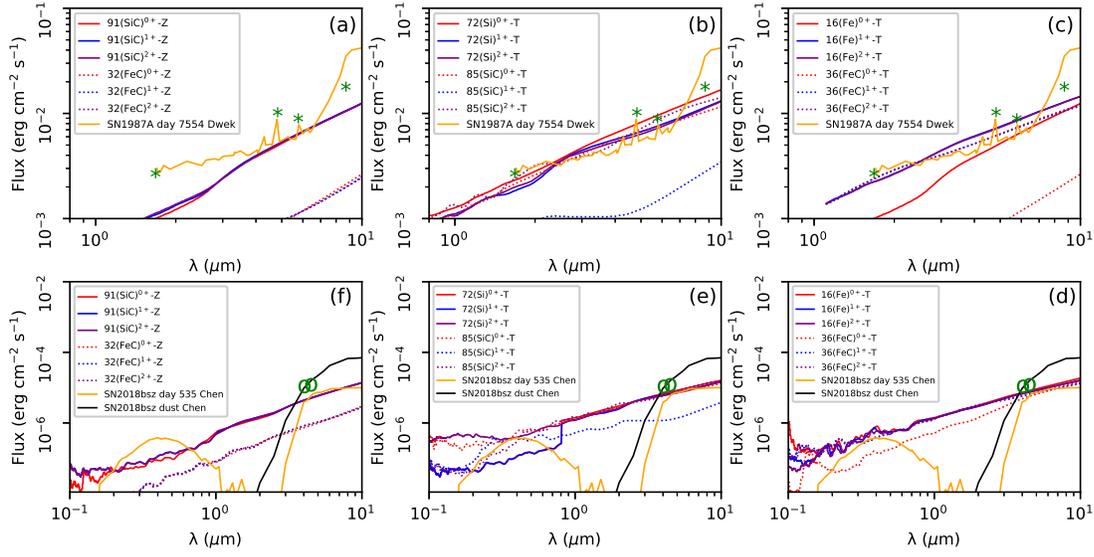}%"scale"后的数字为图形的宽度，也可用"width=1.0\columnwidth" 定义
\caption{Due to the ionization in the SNR, the IR spectrum of dust has obvious changes.
We compare the IR spectrum of 0D and 2D dust with the IR spectrum of SN1987A and SN2018bsz in the panels (a)-(c) and (d)-(f), respectively.
The green circle represents the galactic extinction radiation flux observed by SN 2018bsz.
The black solid line is the radiant flux of the new dust 535 days after the SN ejection predicted by \cite{Chen2021}.
} %图题
\label{Fig5}%{}中"fig:example1"为图名，引用时用\ref{fig:example1}
\end{figure}

\begin{figure}[ht!]%"[]"中为位置参数，四个参数tbph依次是置顶、置底、浮动、当前位置，，选用的参数优先顺序为h-t-b-p
\centering
\includegraphics[scale=0.76, angle=0]{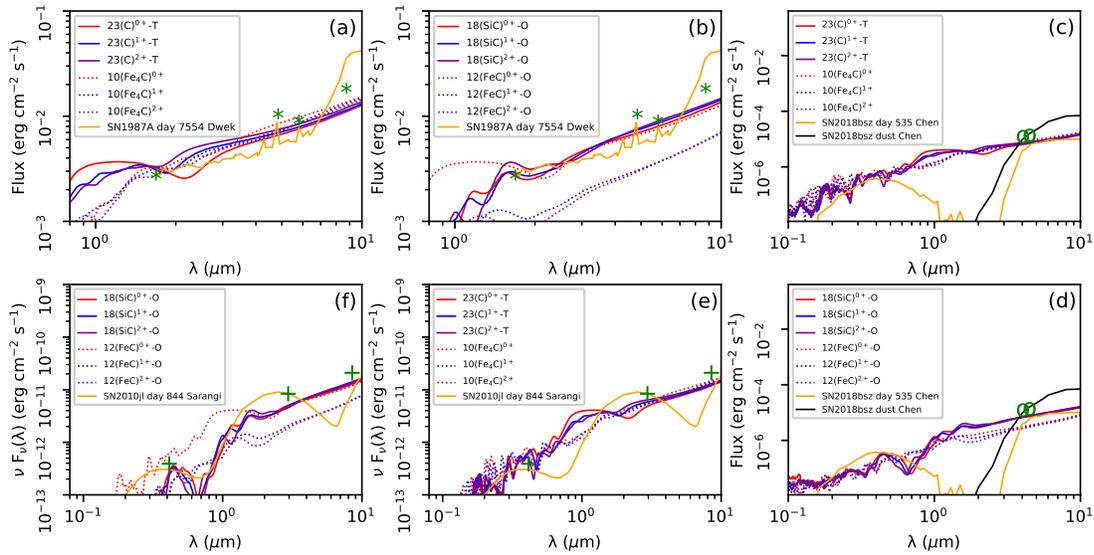}%"scale"后的数字为图形的宽度，也可用"width=1.0\columnwidth" 定义
\caption{Similar to Figure~\ref{Fig5}, in the case of ionization,
the radiation spectra of 1D/2D C$-$rich dust structures are compared with the radiation spectra of SN1987A,
SN2010jl, and SN 2018bsz, respectively.} % 图题
\label{Fig6}%{}中"fig:example1"为图名，引用时用\ref{fig:example1}
\end{figure}

\begin{figure}[ht!]%"[]"中为位置参数，四个参数tbph依次是置顶、置底、浮动、当前位置，，选用的参数优先顺序为h-t-b-p
\centering
\includegraphics[scale=0.76, angle=0]{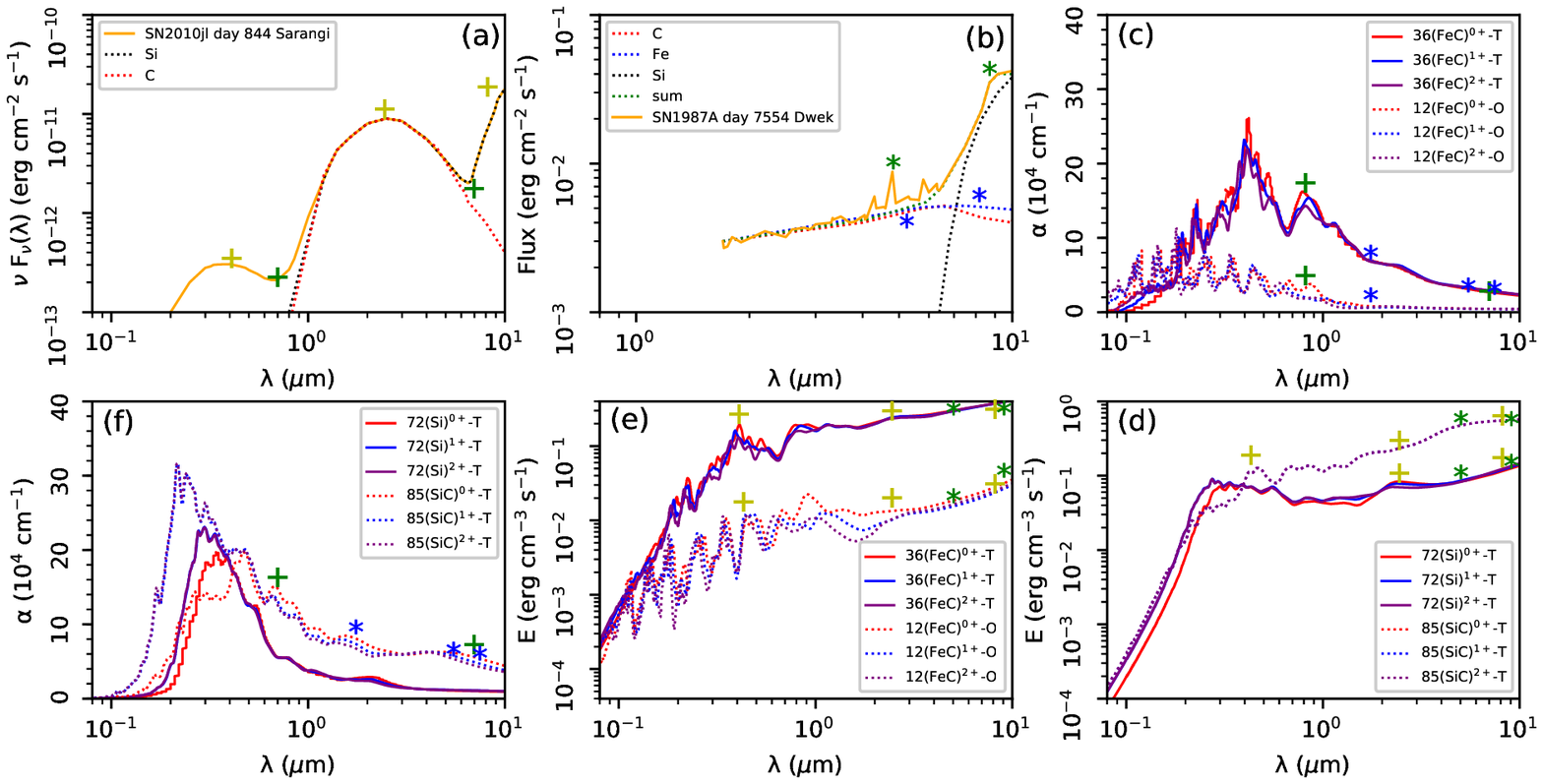}%"scale"后的数字为图形的宽度，也可用"width=1.0\columnwidth" 定义
\caption{Subgraphs (c) and (f) show the light absorption spectra of the 2D structure (36FeC-T, 72Si-T, 85SiC-T) and the
1D structure (12FeC-T) in the ionized state from 0 to +2 charge,
while subgraphs (d) and (e) correspond to the emission coefficients of the above dust structure.
At the same time, it also has corresponding peak value with the SN spectrum (SN2010jl and SN1987A).} % 图题
\label{Fig7}%{}中"fig:example1"为图名，引用时用\ref{fig:example1}
\end{figure}

Figure~\ref{Fig5} and ~\ref{Fig6} show the radiation spectra of the 1D and 2D dust with +1 or +2 charge states.
In the legend of Figure~\ref{Fig5} (a), 91(SiC)$^{0+}$-Z is an electrically neutral dust, 91(SiC)$^{1+}$-Z and 91(SiC)$^{2+}$-Z
are dust structures that lose one and two electrons after ionization, respectively.
We find that most of the structure is basically consistent,
except that the structure of 32FeC-Z is 1 order of magnitude lower than the observed results.
The spectrum of the 1D structure matches the SN2010jl spectrum,
and the 2D dust spectrum fits well with the SN1987A observation data.
But the radiation spectrum of 1D and 2D C$-$rich dust are basically consistent
with the SN2018bsz spectrum when the band range are (0.3-0.7) $\mu$$m$ and (3.5-10) $\mu$$m$.
Tne spectrum of SN2018bsz also shows that C$\footnotesize\textrm{\uppercase\expandafter{\romannumeral2}}$ emission lines are observed
at wavelengths of 0.59, 0.66, and 0.72 $\mu$$m$, respectively \citep{Anderson2018}.
In the panels (d)-(f) of Figure~\ref{Fig5}, in our work,
the size of the dust is the initial size of the model (0.1-0.8) ${\rm \AA}$.
However, the dust size given by \cite{Chen2021} is 0.1 $\mu$$m$.
Therefore, we believe that there may be nano$-$scale C$-$rich dust in the remnants of SN2018bsz.
The results of the SiC dust structure are in good agreement with the observations,
the difference between the neutral and the charged states is minor.
One reason could be that the Fe-rich dust is more active than the Si-rich dust, and that the interaction between particles is stronger,
which leads to easier electron loss and enhances emission coefficient, but its absorption coefficient increases more strongly,
and eventually lead to a lower IR spectrum.
In contrast, both of the 2D dusts 16Fe-T and 36Fe-T with charge of plus one and plus two have higher radiation flux than the electrically neutral dusts,
but the 1D 12FeC-O structure is just the opposite.
In the panel (c) and (e) of Figure~\ref{Fig7},
the light absorption coefficient and emission coefficient of the 2D structure
are higher than those of the 1D structures.
The larger the surface area of the 2D structure, and the greater the probability of photons being emitted,
lead to flux increase in the final IR spectrum.
Moreover, in the panel (e) of Figure~\ref{Fig7}, it is seen that the emission coefficients of the
36FeC-T structure with a charge of 0, plus one and plus two basically remain almost unchanged.
However, in the panel (c) of Figure~\ref{Fig7},
the absorption coefficient of the structure is broadened,
and the 36FeC-T dust has a lower absorption coefficient with a charge of +2.
\textbf{Equation (8)} indicates that the radiation flux and the absorption coefficient are inversely proportional
for a constant the emission coefficient remains constant.
With the increasing of the number of dust charges in the ionized state,
the 1D 12FeC-O dust emission coefficient decreases faster than the absorption coefficient,
which ultimately leads to a decrease in radiant flux.

\section{Conclusions}\label{sec:4}

In this paper, we obtained the IR spectra of C$-$rich dust both in neutral and ionized states through DFT calculations.
Those spectra are compared with the SN observed IR spectrum in the band (0.08-10 $\mu$$m$).

First of all, our results indicate that C$-$rich dust with 2D layered structure has higher radiation flux than those of 0D and 1D structures.
That is the IR spectrum with a larger side length ($\sim$ 72 nm) of dust has higher radiant flux,
while a smaller side length ($\sim$ 9 nm) is just the opposite.
In the remnants of SN2018bsz, there is most likely the dust of nano$-$scale C$-$rich ($\sim$ 0.1 \emph{nm}).
Secondly, we propose that the remnant of SN1987A may has a 2D or 1D structure of the second dual-component dusts SiC or FeC.
Thirdly, 1D structure of the second dual-component dust SiC or FeC, 2D structure single-component dust 23C-T,
and small clusters of dust 10(Fe$_{4}$C) may exist in the SN1987A and SN2010jl.
The spectrum of C$-$rich dust is more consistent with the spectrum of SN2018bsz
both in the band (0.3-0.7) $\mu$$m$ and (3.5-10) $\mu$$m$.

At last, for the dust in ionized states, 2D dusts with +2 charge have higher radiation flux than the electrically neutral dust.
On the contrary, the 1D 12FeC-O structure of the dust is just the opposite.
Our results would contribute to the more precise size and optical properties of C$-$rich dust,
and provide theoretical support for later dust observations in the SNR.
This calculation is of great significance for the future budget of the output of stellar dust,
galaxy dust, and the influence of dust extinction on observations.
For example, it can serve as a reference for dust observations by the James Webb Space Telescope
(JWST), the Large Sky Area Multi$-$Object Fiber Spectroscopy Telescope (LAMOST),
and the upcoming China Space Station Optical Survey Telescope (CSST).

\normalem
\begin{acknowledgements}
We are very grateful to professor ZhanWen Han and Zhengwei Liu from Yunnan Astronomical Observatory of the Chinese Academy of Sciences (YNAO)
for their contributions and help in this article.
We are grateful for resources from the High Performance Computing Center of Central South University.
This work received the generous support of the Independent Innovation Project for Postgraduates of
Central South University No, 160171008.
The National Natural Science Foundation of China, projects No, 11763007, 11863005, 11803026, and U2031204.
In the end, we would also like to express our gratitude to the Natural Science Foundation of Xinjiang No.2021D01C075.

\end{acknowledgements}

\bibliographystyle{raa}
\bibliography{bibtex}

\end{document}